\documentclass[prl,twocolumn]{revtex4}
\usepackage{epsfig}
\usepackage{amsmath}
\usepackage{epsfig}

\begin{document}
\title
{Divergence of $<\!p^6\!>$ in discontinuous potential wells}    
\author{Zafar Ahmed$^1$, Sachin Kumar$^2$, Dona Ghosh$^3$, Joseph Amal Nathan$^4$}
\affiliation{$^1$Nuclear Physics Division, $^2$Theoretical Physics Section, $^4$Reactor Physics Design Division, Bhabha Atomic Research Centre, Mumbai 400085, India\\
	$^3$Department of Mathematics, \nopagebreak Jadavpur University, Jadavpur, Kolkata, 700032, India}
\email{1:zahmed@barc.gov.in,   2:sachinv@barc.gov.in, 3:rimidonaghosh@gmail.com, 4:josephan@barc.gov.in}
\date{\today}
\begin{abstract}
The surprising divergence of the expectation value 	$<\!p^6\!>$ for the square well potential is known. Here, we prove and demonstrate the  divergence of $<\!p^6\!>$ in potential wells which have a finite jump discontinuity; apart from the square-well two-piece half-potentials wells are examples. These half-potential wells can be expressed as $V(x)=-U(x) \Theta(x)$, where $\Theta(x)$ is the Heaviside step function. $U(x)$ are continuous and differentiable functions with  minimum at $x=0$ and which may or not vanish as $x\sim \infty$. 
\end{abstract}
\maketitle

In quantum mechanics [1-3] for a potential well the expectation value $<\!\psi(x)|F|\psi(x)\!>$ of an operator $F$ is obtained using  eigenfunction of a bound state that  is a continuous and normalizable solution of Schr{\"o}dinger equation
\begin{equation}
\frac{d^2\psi(x)}{dx^2}+\frac{2m}{\hbar^2}[E-V(x)] \psi(x)=0.
\end{equation}
Recently, it has been pointed out that $\psi(x)$ needs to vanish faster than $|x|^{-3/2}$ [4] additionally in order to have a finite value for $<\! x^2 \!>$ and  the uncertainty in position $\Delta x$. Otherwise, the state will be bound but (infinitely) extended state.  Interestingly, the potential $V(x)$ having ground state where the asymptotic fall-off is slower than this are found to have only one bound state.

One is also advised to work in momentum representation [1-3] where the wave function $\phi(p)$ is given as
\begin{equation}
\phi(p)={\cal F}[\psi(x)]=(2\pi \hbar)^{-1/2} \int_{-\infty}^{\infty} \psi(x) e^{-ipx/\hbar} dx,
\end{equation}
the Fourier transform of  $\psi(x)$: ${\cal F}$$[\psi(x)]$. The two representations are physically equivalent.
One can find $<\!x^2\!>$ as $<\!\psi(x)|x^2|\psi(x)\!>$ or $<\!\phi(p)|-\hbar^2(d^2/dp^2)|\phi(p)\!>$. Similarly, $<p^2>$ can be found
as $<\!\psi(x)|-\hbar^2(d^2/dx^2)|\psi(x)\!>$ or $<\!\phi(p)|p^2|\phi(p)\!>$. We can again demand that in order to have $<\!\phi(p)|p^2|\phi(p)\!>$ and $\Delta p$ as finite, $\phi(p)$ needs to vanish faster than $|p|^{-3/2}$. The question arising is as to what property of $V(x)$  ensures a finite value for $<\! p^2 \!>$ in a potential well.

Most often the mathematical forms of $\psi(x)$
and $\phi(p)$ are different so much so that for finding something, one option is either easier to do or more transparent than the other one.
Also these two options present different mathematical situations. For instance, for infinite square well (ISW),  the $p$-integral
in finding $<p^2>$ is improper [5] whereas the $x$-integrals are proper and simple. For ISW, $<\!p^4\!>$ in the position representation gives a finite
value, it actually diverges in momentum space. Similar experience is found [6] in finite square well (FSW) where it is $<\! p^6 \!>$ which presents an interesting  discrepancy in the two representations.

This discrepancy was first pointed out in a largely un-noticed paper [7] where for FSW $|\phi(p)|^2$ was derived to show a surprising asymptotic fall-off as $p^{-6}$, however the details of $\phi(p)$ were incorrect which have been corrected recently [6]. The consequent  divergence of $<\! p^6 \!>$ in FSW in position space  was revealed in terms of the Dirac delta discontinuities in the second and higher order derivatives of $\psi(x)$ at the end points $x=\pm a$. Unfortunately, this proof [7] turns out to be  tautological even for FSW.  

Here, in this paper, we wish to prove and demonstrate the general divergence of $<\!p^6\!>$ when a potential well has a finite jump-discontinuity. Apart from square well, two-piece half-potentials wells (Fig. 1) like
\begin{equation}
V(x)=-U(x) \Theta(x), \quad \Theta(x<0)=0, \Theta(x\ge 0)=1.
\end{equation}
where $U(x)$ are continuous and differentiable  functions with non-zero  minimum at the junction ($x=0$). $U(x)$ may or not vanish as $x \rightarrow \infty$, see  Fig. 1. We also call them as half-potential wells because for $x > 0$ they are half parts of the well known parabolic $[U(x)=-V_0(1-x^2/a^2)]$ [1], triangular $[U(x)=-V_0(1-|x|/a)]$ [2], Eckart $[U(x)=-V_0 \mbox{sech}^2(x/a)]$ [2,3] and exponential $[U(x)=-V_0(2-\exp(-2|x|/a)]$ [9] wells. 

\begin{figure}
	\centering
	\includegraphics[width=8 cm,height=5 cm]{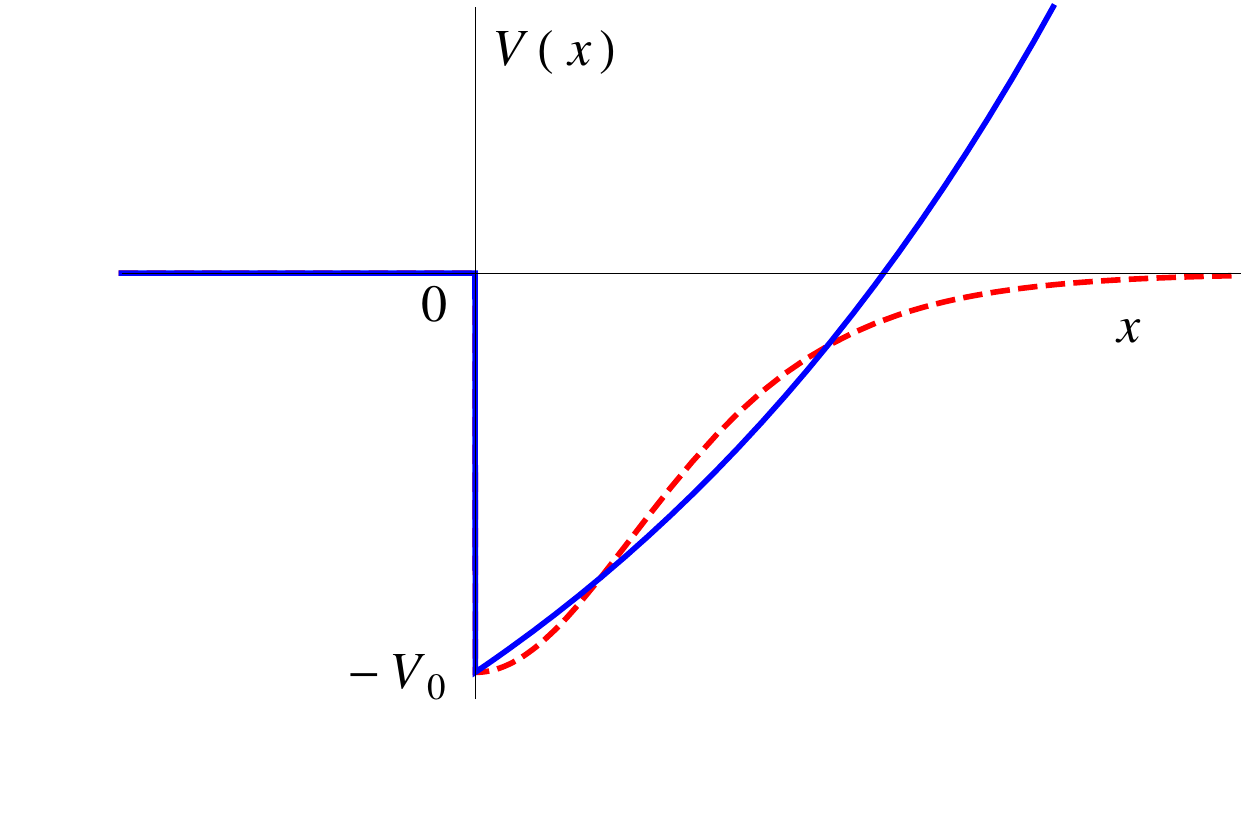} 
	\caption{Schematic representation of  discontinuous two-piece half potential wells  (3). The solid one represents
	the parabolic (17), the triangular (21) and the exponential (29) models which diverge asymptotically. The dashed curve represents the Eckart (25) model which converges asymptotically.}
\end{figure}

A  definite integral $\int_{a}^{b} f(x) dx$ is real and finite if it is continuous at each and every point of the domain $[a,b]$, $f(x)$ may also be piece-wise continuous for this integral to exist. Otherwise, the integrals are improper which may be convergent (finite) or divergent (infinite) [5]. One can evaluate the expectation value of $p^2$ for $n^{th}$ bound state as
\begin{equation}
<\!\psi_n|p^2|\psi_n\!>  =\frac{2m}{\hbar^2} \int_{-\infty}^{\infty} \psi_n(x) [E_n-V(x)]~ \psi_n(x) dx,
\end{equation}
which is easily finite. Let us fix $2m=1=\hbar^2$ for  the sequel. The eigenstate $\psi_n(x)$  being both continuous and differentiable in $(-\infty, \infty)$, for continuous $V(x)$ the above integral is finite. $V(x)=-2\delta(x)$ is an  interesting digression 
where in $<\!p^2\!>$ is finite owing to the interesting property that $\int_{-\infty}^{\infty} f(x) \delta(x) dx =f(0).$ 

Next, we suggest $<\! p^4 \!>$ to be evaluated  as 
\begin{equation}
- \int_{-\infty}^{\infty} \psi(x) \frac{d^2}{dx^2}[E-V(x)] \psi(x) dx,
\end{equation}
which can be re-written in an inspiring form as
\begin{multline}
\hspace*{-0.5 cm}{<}\!p^4\!{>}{=}{\int_{-\infty}^{\infty}}{F[\psi,\psi',V,{V'},V{''}]} dx{+}{<}\psi{|(E{-}V(x))^2|}\psi{>}. \hspace{-.15cm}
\end{multline}

Ordinarily, the first integral in above simplifies to $[-V'(x) \psi^2(x)]_{-\infty}^{\infty}$. When $V(x)$ is continuous and differentiable, it vanishes since $\psi(x)$ are bound states
that converge to zero, asymtotically. Alternatively,  inside the first integral in (6),  there occur  terms like $-2[V'(x) \psi(x) \psi'(x)+V''(x) \psi^2(x)]$. For the Dirac delta well $V(x)=-2\delta(x)$ using the interesting derivatives   $V'(x)=2\delta(x)/x$ and $V''=-4\delta(x)/x^2$; $\psi_0(x)=e^{-|x|}$, the second term causes strong divergence in $<\!p^4\!>$ near $x=0$ as
\begin{equation}
\int_{-\epsilon}^{\epsilon} V''(x) \psi^2(x) dx=2 \int_{-\epsilon}^{\epsilon} e^{-2|x|} \frac{\delta(x)}{x^2} dx \rightarrow \infty.
\end{equation}
Had there been odd eigenstate(s) this integral would have been convergent and finite. This explains the divergence of $<\! p^4 \!>$ in position space which is obvious in momentum space as $\phi(p)=\sqrt{2/\pi} (1+p^2)^{-1} [8].$ 
Next, we verify that expectation value of force $(-V'(x))$, namely
\begin{equation}
\hspace*{-0.5 cm}{<}\!\psi{|V'(x)|}\psi\!{>}{\rightarrow }2\int_{{-}\epsilon}^{\epsilon}\frac{\delta(x)}{x}\psi^2(x)dx{\rightarrow} 2\int_{{-}\epsilon}^{\epsilon} \frac{\delta(x)}{x}(1{-}2{|}x{|}) dx 
\end{equation}
vanishes as here is an odd integrand between symmetric limits. Vanishing of integrals in (8), may not be without arguments. Here, we underline that otherwise Ehrenfest theorem will be defied by
the Dirac delta well potential which is most popular among potential wells. 

By virtue of the fact that eigenfunctions of bound states are continuous and differentiable at any point in the domain of the potential one can plot a tangent on $\psi(x)$ that behaves locally as
\begin{equation}
 \psi(x) \approx \alpha+\beta (x-x_0), \forall ~ x_0 \in (-\infty, \infty).
\end{equation} 
in the close  vicinity of any point $x=x_0$. Specially, for half-potential wells discussed here due to asymmetry of the potential, even around $x=0$,  $\alpha \ne 0$.
Strangely, in Ref. [7] (see above Eq. (17) there), $\psi(x)$ near the point of discontinuity of FSW  has been assumed to be $\psi(x) \approx \gamma (x-x_0)^2$  which cannot be true, since  a more general approximation could be as  $\psi(x) \approx \alpha+ \beta (x-x_0)+ \gamma (x-x_0)^2$, where in any case the linear term will dominate by orders.
This makes his proof of divergence of $<\! p^6 \!>$ in FSW as tautological.

First, we would like to  give  a correct proof of convergence of $<\!p^4\!>$ and the divergence of $<\!p^6\!>$ in the square well potential [7]
\begin{equation}
V(x)=-V_0[\Theta(x)-\Theta(x-a)].
\end{equation}
Let us consider the integrals of expectation values (6,10) in the infinitesimal domain $(-\epsilon, \epsilon)$ around $x=0$. Here we have $V'=-V_0[\delta(x)-\delta(x-a)], V''=-V_0 [\delta'(x)-\delta'(x-a)], V'''=-V_0 [\delta''(x)-\delta''(x-a)],
V^{iv}(x)=-V_0 [\delta '''(x)-\delta'''(x-a)]$. As seen above in (7) that $<\psi(x)|V''(x)|\psi(x)>$ is the main source of divergence in $<\! p^4 \!>$. So for the square well we can write
\begin{multline}
<\psi| V^{''}|\psi>=-V_0\int_{-\infty}^{\infty} [\delta'(x)-\delta'(x-a)] \psi^2(x) dx \\ \rightarrow V_0 \left(\alpha^2 \int_{-\epsilon}^{\epsilon} \frac{\delta(x)}{x} dx-  \alpha_1^2 \int_{-\epsilon}^{\epsilon} \frac{\delta(t)}{t} dt \right)
\end{multline}
$\alpha$ and $\alpha_1$ are due to local behavior of $\psi(x)$ as per (9) at point $x=0$ and $x=a$, respectively. As done in Eq. (8), these two integrals vanish and hence  $<\! p^4\!>$ is convergent.
In $<\!p^6\!>$ the source of divergence is the term $<\! V^{iv}(x)\!>$, which in the infinitesimal
domain around $x=0$ and $x=a$ can be written as
\begin{multline}
<\psi| V^{iv}| \psi>=-V_0\int_{-\infty}^{\infty} [\delta'''(x)-\delta'''(x-a)] \psi^2(x) dx \rightarrow \\ 
-2V_0 \left ( \alpha \beta  \int_{-\epsilon}^{\epsilon} \frac{\delta(x)}{x^2} dx- \alpha_1 \beta_1 \int_{-\epsilon}^{\epsilon} \frac{\delta(t)}{t^2} dt \right ).
\end{multline}
As in Eq. (7) both of these integrals are divergent and so is $<\! p^6 \!>$ for square well (10).
The expectation value of force for (10) is
\begin{equation}
<\! \psi_n(x)|V'(x)|\psi_n(x)\!>=V_0[\psi^2_n(0)-\psi^2_n(a)]=0,
\end{equation}
which can be verified by using the eigenfunctions of square well potential.

Further, owing to interesting derivatives namely $\Theta'(x)=\delta(x)$ and $\delta'(x)=-\delta(x)/x$, for the potentials (3)  one part of the terms  $<\! \psi(x)| V''(x)| \psi(x) \!>$ in (6)  in the infinitesimally small domain $(-\epsilon, \epsilon)$ appears as
\begin{equation}
-\int_{-\epsilon}^{\epsilon}  U(x) \frac{\delta(x)}{x}  (\alpha+\beta x)^2  dx\rightarrow V_0 \alpha^2 \int_{-\epsilon}^{\epsilon} \frac{\delta(x)}{x}~ dx,
\end{equation}
which can be taken to vanish as in Eq.(8). Therefore, $<\!p^4\!>$ for the type of potentials (3) discussed here is convergent.

For the expectation value of $<\!p^6\!>$, we get 
\begin{multline}
\hspace{-.35cm}<\!p^6\!>{=}\int_{{-}\infty}^{\infty} {F_2[}\psi,\psi',\psi'',\psi''',\psi^{(4)},V,{V'},{V''},{V'''},V^{(iv)}]\\dx{+}<\!\psi|[E-V(x)]^3|\psi \!>.
\end{multline}
In the above equation the part $<\! \psi(x)|V^{iv}(x)| \psi(x)\!>$ is the main source of divergence in $<\!p^6\!>$.

The potentials (3) which are piece-wise continuous with finite jump discontinuity at $x=0$, the successive derivatives of $V(x)$ are:  $V'(x)=U'(x) \Theta(x)+U(x) \delta(x)$. $V''(x)=U''(x)\Theta(x)+2 U'(x) \delta(x)+U(x)\delta'(x)$, $v'''(x)=U'''(x)+3U''(x)\delta(x)+3U'(x)\delta'(x)+U(x) \delta''(x)$  and $U^{iv}(x) \Theta(x)+4U'''(x)\delta(x)+6U''(x)\delta'(x)+4U'(x)\delta''(x)+U(x)\delta'''(x).$
Noting some very interesting derivatives as $\Theta'(x)=\delta(x)$, $x\delta'(x)=-\delta(x)$,
$x\delta^{''}(x)=2\delta(x)/x$, $x\delta^{'''}(x)=-6\delta(x)/x^2$, we find that in $<\!p^6\!>$ (10)
the term $<\psi(x)|V^{iv}(x)|\psi(x)>$ can cause divergence in the infinitesimal domain $(-\epsilon, \epsilon)$ as
\begin{multline}
\hspace{-.3 cm}I{=}\int_{{-}\epsilon}^{\epsilon}U(x){\delta'''}(x)~\psi^2(x)~dx \rightarrow 12 \alpha \beta V_0 \int_{{-}\epsilon}^{\epsilon} \frac{\delta(x)}{x^2}dx, 
\end{multline}
which diverges.

We would like to re-emphasize that in our proofs given above
our assumption for the eigenstates (9) and vanishing of the integral $\int_{-\epsilon}^{\epsilon} [\delta(x)/x] dx$, play the most crucial role. 

In the following, we present four analytically solvable models of Half-potential wells (3) wherein much longer tail of $p^6 I(p)$ would justify the acclaimed (12)  divergence of $<\!p^6\!>$. $I(p)$ denotes the momentum distribution calculated as $I(p)=|\phi(p)|^2$, where $\phi(p$) is the Fourier transform (2) of $\psi(x)$ that will be obtained in the sequel. In the fifth and the sixth  solvable models which are continuous but non-differentiable wells,  we show that $<\!p^6\!>$ is convergent.
\vskip .5 cm
{\bf 1. Half-parabolic well:}
This potential is written as 
\begin{equation}
V(x)=-V_0 \left[1-\frac{x^2}{a^2}\right] \Theta(x),
\end{equation}
For $x\ge 0$, the Schr{\"o}dinger equation (1) for (17)  can be transformed to parabolic cylindrical differential equation  [1,10] as
\begin{multline}
\frac{d^2\psi}{dz^2}{+}\left[\nu{+}\frac{1}{2}{-}\frac{z^2}{4}\right] \psi{=}0, \quad z{=}\gamma x,  \quad \nu{=}\frac{E{+}V_0}{\omega}{-}\frac{1}{2}.
\end{multline}
where $\omega=\sqrt{\frac{2V_0 \hbar^2}{ma^2}}$, $\gamma= \sqrt[4]{ \frac{8mV_0}{\hbar^2 a^2}}$.
One of the two linearly independent solutions $D_{\nu}(\pm z)$ of this equation is $D_{\nu}(z)$ which vanishes at $z=\infty$ can be taken to be correct solution for $x\ge0$ as $\psi(x\ge 0)= B D_{\nu}(z)$. The solution of (1) for $x<0$, is sought as $\psi(x<0)= A e^{kx}$. By matching these two pieces and their derivative at $x=0$, we get eigenvalue equation as
\begin{equation}
\gamma D_{\nu}'(0)= k D_{\nu}(0),\quad k=\sqrt{\frac{-2mE}{\hbar^2}}. 
\end{equation}
The energy eigenfunctions are given as 
\begin{equation}
\psi(x<0)= C~ D_{\nu}(0)~ e^{kx}, \psi(x\ge 0)= C~ D_{\nu}(\gamma x). 
\end{equation}
We propose to fix $2m=1= \hbar^2$, $a=2, V_0=15$ in arbitrary units for all the wells to be considered in the sequel. Solving (19) we find bound states  at $E=-10.6370, -3.9894$. In Fig 2, we plot the distributions $p^2 I(p)$ (dotted line), $p^4 I(p)$ (dashed line) and $p^6 I(p)$ (solid line) for the ground state.  Notice the much longer  tail in the solid curve than that of the dashed curve indicating that $<\! p^6 \!>$ would actually diverge.
\begin{figure}
	\centering
	\includegraphics[width=8 cm,height=5 cm]{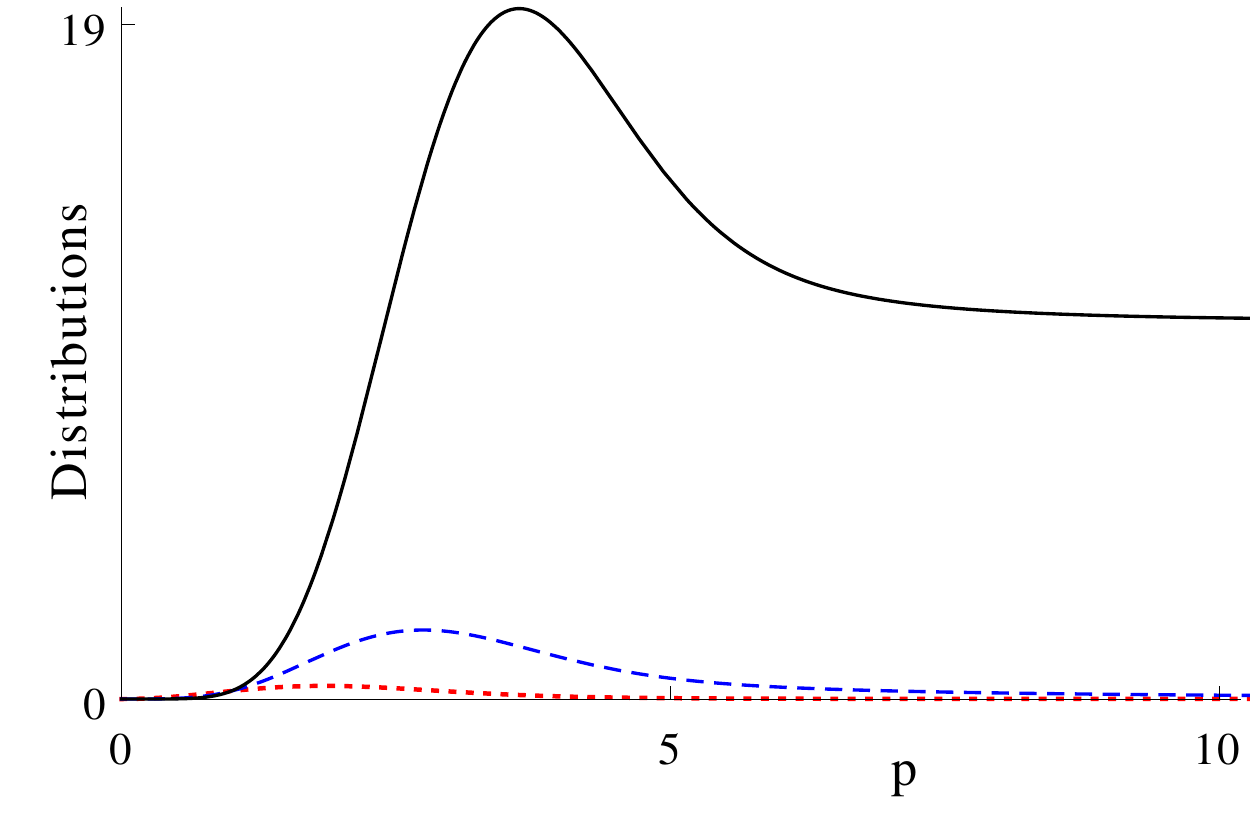}
	\caption{The momentum distributions $p^{2j} I(p)$, $j=1$: dotted line, $j=2$: dashed line, $j=3$: solid line for the ground state of the half-parabolic potential (17) when $V_0=15, a=2$ and $2m=1= \hbar^2$. Notice the  much longer tail in the solid curve indicating the divergence of $<\!p^6\!>$.}
\end{figure}
\vskip .5 cm
{\bf 2. Half-triangular well:} This potential is given as
\begin{equation}
V(x \ge 0)=-V_0 \left[1-\frac{x}{a}\right], \quad V(x > 0)=0.
\end{equation}
The Schr{\"o}dinger equation (1) for this potential when $x\ge 0$ can be transformed to the Airy differential equation [2,10] as
\begin{multline}
\hspace*{-0.5 cm}\frac{d^2\psi}{dy^2}{-}y \psi{=}0, y(x){=}\frac{2m}{g^2\hbar^2}\left[\frac{V_0x}{a}{-}E{-}V_0\right], g{=}\sqrt[3]{\frac{2mV_0}{\hbar^2 a}}.
\end{multline}
This second order equation has two linearly independent solutions called Airy functions $Ai(y)$ and $Bi(y)$. It is $Ai(y)$ that vanishes as $x \sim \infty$, so we admit  the solution of  (22)
as $\psi(x\ge 0) = B Ai(y)$ and for $x<0$, we have $\psi(x < 0)= A e^{kx}$. Matching these two solutions and their derivative at $x=0$, we obtain the energy quantization condition as
\begin{equation}
g Ai'(y_0)=k Ai(y_0), \quad  y_0=-\frac{2m}{\hbar^2}\frac{E+V_0}{g^2}. 
\end{equation}
For these discrete energies the eigenfunctions are given as
\begin{equation}
\psi(x<0)=C Ai(y_0) e^{kx},\quad \psi(x\ge 0)= C Ai(y(x)).
\end{equation}
We take $V_0=15$ and $a=2$ in arbitrary units, the well has two bound states at  $E=-8.1408$ and -1.8025. For the ground state, we plot various distributions as in Fig. 3. We confirm the long tail
in $p^6 I(p)$ that would give rise to divergence in $<\! p^6 \!>$.
\begin{figure}
	\centering
	\includegraphics[width=8 cm,height=5 cm]{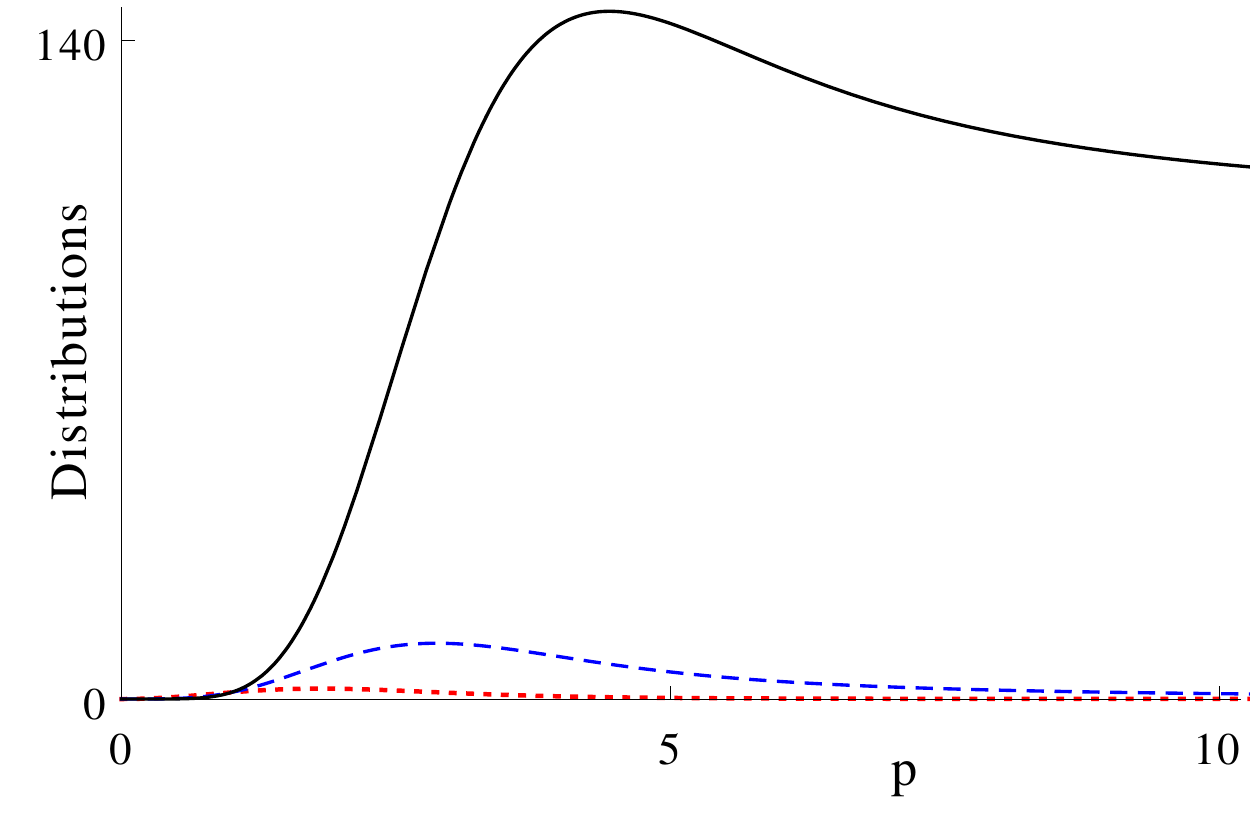}
	\caption{The same as in Fig. 2, for the half-triangular potential (21). Notice the  much longer tail in the solid curve indicating the divergence of $<\! p^6 \!>$.}
\end{figure}
\vskip .5 cm
{\bf 3. Half-Eckart potential:}
This potential is expressed as 
\begin{equation}
V(x\ge 0)=-V_0 \mbox{sech}^2 (x/a), V(x < 0)=0
\end{equation}
The Schr{\"o}dinger equation (1) for this potential when $x \ge 0$ can be transformed  to the Gauss hypergeometric equation as [2,3,10]
\begin{multline}
z(1{-}z)F''{+}(ka{+}1)(1{-}2z)F'{-}(ka{-}s)(ka{+}s{+}1)F=0,\\ 
z{=} \frac{1{-}\tanh(x/a)}{2}, ~~ s{=}\frac{1}{2} \left({-}1{+}\sqrt{\frac{8mV_0 a^2}{\hbar^2}}\right)
\end{multline}
The solution of (26) which vanish for $x \sim \infty$ is given as
\begin{equation}
\psi(x{\ge}0){=}C(1{-}z^2)^{\frac{ka}{2}}~_2F_1[ka{-}s,ka{+}s{+}1,ka{+}1,\frac{1}{2}(1{-}z)].
\end{equation}
For $x<0$, we have $\psi(x<0)=C _2F_1[ka-s,ka+s+1,ka+1;1/2] e^{kx}$, by matching these two solutions and their derivative at $x=0$,
we get the eigenvalue equation in terms of hypergeometric function $~2F_1[\boldsymbol{a},\boldsymbol{b},\boldsymbol{c};1/2]$ [10] which are known it terms of Gamma functions $\Gamma[t]$ [10]. Utilizing such results, we obtain the eigenvalue equation as
\begin{equation}
ka+2\frac{\Gamma[(1+ka-s)/2] \Gamma[(2+ka+s)/2]}{\Gamma[(ka-s)/2]\Gamma[(1+ka+s)/2]}=0.
\end{equation} 
For $V_0=15$ and $a=2$, we get four bound states in the potential at $E=-10.9628,-5.8470,-2.2641,-0.3400$. Distributions: $p^{2j} I(p),j=1,2,3$ for ground state are shown in Fig. 4.
\begin{figure}
	\centering
	\includegraphics[width=8 cm,height=5 cm]{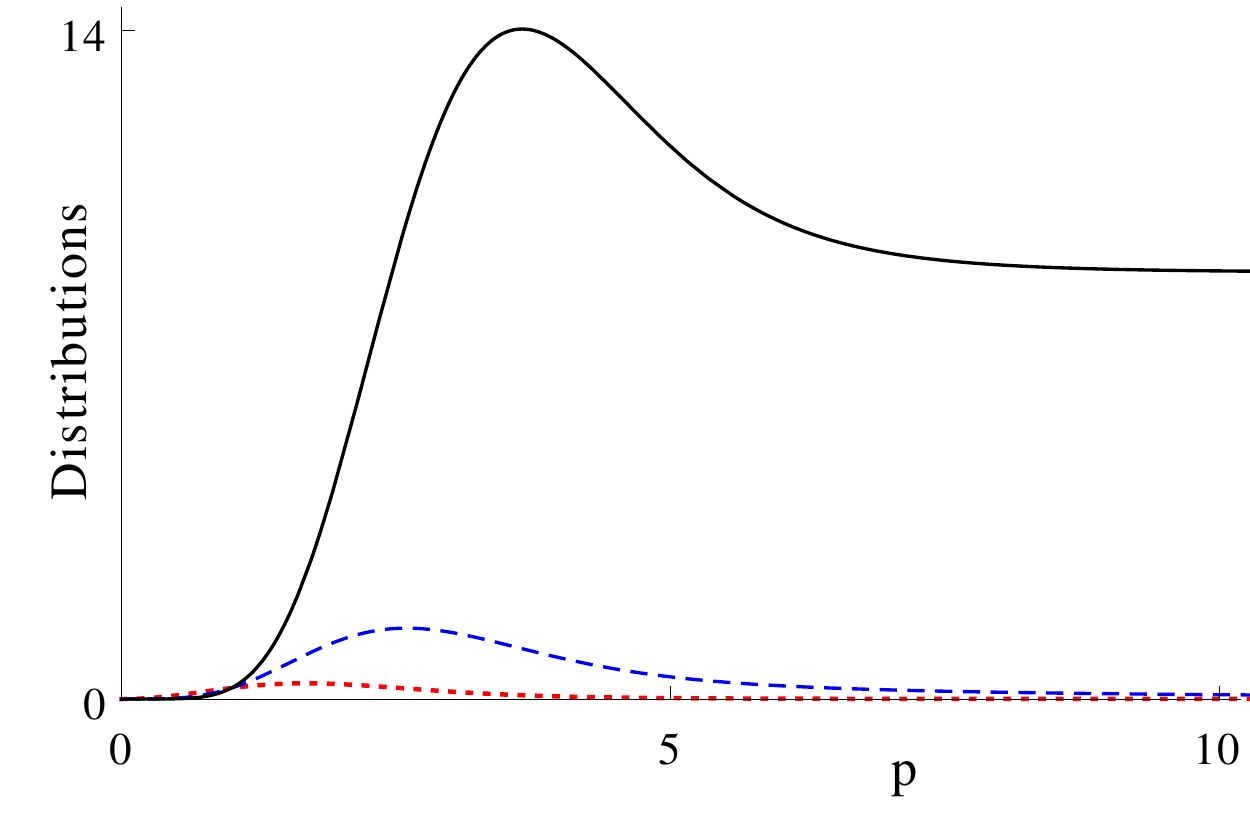}
	\caption{The same as in Fig. 2, for the half-Eckart potential (25). Notice the  much longer tail in the solid curve indicating the divergence of $<\! p^6 \!>$.}
\end{figure}
\vskip .5 cm
{\bf 4. Half-exponential well:} This potential is written as
\begin{equation}
V(x \ge 0)=-V_0 [2-e^{2x/a}],~~ V(x < 0)=0
\end{equation}
for which the Schr{\"o}dinger equation (1) when $x \ge 0$ can be transformed to the cylindrical Bessel equation
as [9,10]
\begin{eqnarray}
z^2\frac{d^2\psi}{dz^2}+ z\frac{d\psi}{dz}+(-\kappa^2 a^2-z^2) \psi=0,\quad  z=qae^{x/a}, \nonumber \\ \kappa=\frac{\sqrt{2m(E+2V_0)}}{\hbar},  q=\frac{\sqrt{2mV_0}}{\hbar}.
\end{eqnarray}
Out of two linearly independent solutions of (30) as modified Bessel function: $I_{\nu}(z)$ and $K_{\nu}(z)$. Here, we choose $K_{i\kappa a}(z)$ as the solution of (26) for $x > 0$ since it vanishes for $x\sim \infty$. On the other hand, we have  $\psi(x < 0) =A e^{kx}$. We match these two solutions and their derivative at $x=0$ to get the energy eigenvalue equation
\begin{equation}
q K'_{i\kappa}(qa)-k K_{i\kappa a}(qa)=0.
\end{equation} 
For the discrete energy roots of this equation we get the eigenfunctions for (29) as
\begin{equation}
\hspace*{-0.25 cm}\psi(x{<0}){=}CK_{i\kappa a}(qa)e^{kx},~\psi(x \ge 0){=}C K_{i\kappa a}(qae^{x/a}).
\end{equation}
For $V_0=15$ and $a=2$,  we get one bound state in the potential at $E=-3.9249$. The three distributions are plotted in Fig. 5, where the solid line yet again presents much longer tail
justifying the divergence of $<\! p^6 \!>$.
\begin{figure}
	\centering
	\includegraphics[width=8 cm,height=5 cm]{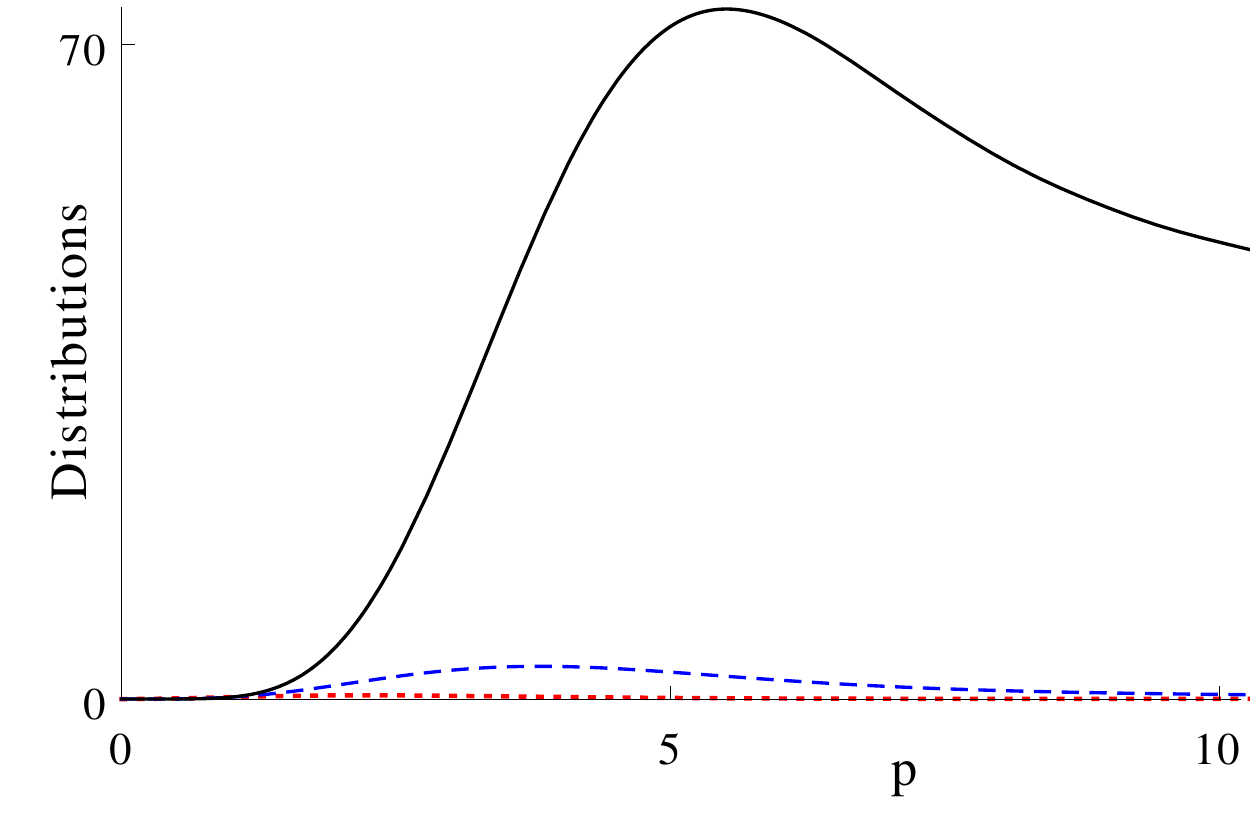}
	\caption{The same as in Fig. 2, for the half-exponential potential (29). Notice the much longer tail in the solid curve indicating the divergence of $<\! p^6 \!>$.}
\end{figure}

Not shown here are the momentum space distributions for excited states, we would like to mention that they too have 
$<\! p^6 \!>$ as divergent and hence $<\! p^{2j}\!>, j=4,5,6..,$ also diverge.  Also, $<\! p^{2j+1} \!>$ for $j=0,1,2,..$ vanish due to antisymmetry of the integrands. Momentum distributions for other interesting one-dimensional potential wells can be seen in Refs. [11,12].
 
 The expectation value of force for delta well has been shown in Eq. (8) to follow Ehrenfest theorem in an interesting way. We have checked that the bound states of  all the half-well potentials (3) discussed here indeed comply to the Ehrenfest theorem as
 \begin{equation}
 \hspace{-.3cm}{<}\!\psi_n{|}V'(x){|}\psi_n(x)\!{>}{=}{-}{\int_{0}^{\infty}}U'(x){\psi_n^2(x)}dx{-}V_0 \psi^2_n(0){=}0,
 \end{equation}
 in  mathematically different ways.

In this paper, we have stated, proved and demonstrated  the divergence of $<\! p^6 \!>$ in potential wells which have a finite jump discontinuity. Our proofs are  in position space and demonstrations in Figs. 2-5 are in momentum space. Our proofs are simple and transparent which also deserve to be the correct proofs for the square well case. We underline the fact that though both representations are physically equivalent yet one is more convenient than the other one for a given purpose. Four analytically solvable models and the much longer tails of their distributions of $p^6 I(p)$ in Figs. 2-5 testify to our claim. Normally, the two piece half-potential wells would be  passed off like other one piece finite potential wells which have finite number of bound states. The present work brings out a distinction between the two. For these half-potential wells since $<\!p^2\!>$ is finite so is the uncertainty product.  So these models would continue to be of interest in normal physical applications. Additionally, now or in future if there is a requirement for the wells which entail divergence of $<\!p^6\!>$ (sixth moment of the momentum distribution), the solvable models discussed here will be up for new considerations. It turns out that just one finite-jump discontinuity in the potential well causes the divergence of $<\!p^{2j}\!>$ for $j=3,4,5,...$ The Dirac delta well turns out to be a unique well where it is $<\!p^4\!>$ which is divergent.

\end{document}